# The LOFT Wide Field Monitor simulator

Immacolata Donnarumma[a], Yuri Evangelista[a,b], Riccardo Campana[a,b], Jean in't Zand[c], Marco Feroci[a,b], Niels Lund[d], Søren Brandt[d], Jörn Wilms[e], Christian Schmid[e]

[a] INAF-IAPS, Via del Fosso del Cavaliere 100, I-00133 Roma, Italy;
[b] INFN-Roma2, via della Ricerca Scientifica 1, I-00133 Roma, Italy;
[c] SRON Netherlands Institute for Space Research, Sorbonnelaan 2, 3584 CA Utrecht, The Netherlands;
[d] Technical University of Denmark – DTU SPACE, National Space Institute, Elektrovej Building 327, DK-2800, Kgs. Lyngby, Denmark;
[e] University of Erlangen-Nuremberg & Erlangen Centre of Astroparticle Physics, Erwin-Rommel-Straße 1 D-91058 Erlangen

## ABSTRACT

We present the simulator we developed for the Wide Field Monitor (WFM) aboard the Large Observatory For X-ray Timing (LOFT) mission, one of the four ESA M3 candidate missions considered for launch in the 2022–2024 timeframe. The WFM is designed to cover a large FoV in the same bandpass as the Large Area Detector (LAD, almost 50% of its accessible sky in the energy range 2–50 keV), in order to trigger follow-up observations with the LAD for the most interesting sources. Moreover, its design would allow to detect transient events with fluxes down to a few mCrab in 1-day exposure, for which good spectral and timing resolution would be also available (about 300 eV FWHM and 10 $\mu$s, respectively). In order to investigate possible WFM configurations satisfying these scientific requirements and assess the instrument performance, an end-to-end WFM simulator has been developed. We can reproduce a typical astrophysical observation, taking into account both mask and detector physical properties. We will discuss the WFM simulator architecture and the derived instrumental response.

## 1. INTRODUCTION

The WFM aboard LOFT (2–80 keV) has been designed to cover a large portion ($\sim 50\%$) of the sky that is accessible to the LAD,[1] which operates in the energy range 2–30 keV (extended energy range 2–80 keV) and is mainly devoted to X-ray timing and spectral studies. Its current configuration matches two important criteria, the overlap of the LAD energy range and the large field of view, in order to trigger follow-up observations for the most interesting sources, both Galactic and extragalactic. Its design is based on the heritage of the SuperAGILE experiment,[2] the hard X-ray monitor aboard the AGILE satellite in orbit since 2007,[3] which demonstrated the feasibility of a compact, large-area, light and low-power, arcminute resolution X-ray imager, with steradian-wide field of view. The employment of the Silicon Drift Detector technology for the LOFT WFM provides a substantial improvement in terms of the lower energy threshold (2 keV) and energy resolutions, with respect to the Si microstrip technology adopted in SuperAGILE. For details of SDD imaging capability we refer to [4] and [5]. It is worth noticing here that SDDs provide an asymmetric spatial resolution, which is < 100 $\mu$m FWHM in the anodic direction (fine direction) and < 8 mm FWHM in the drift direction (coarse direction), achieved by using a one-dimensional read-out system. This means that when combined with a coded mask, each detector is able to provide a 2-dimensional source positioning, with an accuracy that mainly depends on the ratio between the mask element size and the mask-detector distance. However, since the sensitivity is a trade-off between angular resolution and "coding power"[6] a mask element size about two times larger than the worst detector spatial resolution for both fine and coarse directions has been fixed: 250 $\mu$m ×16.125 mm. The resulting angular resolution obtained by positioning a coded mask at $\sim$ 205 mm distance is 4.3 arcminutes and 4.8 degrees in fine and coarse directions, respectively (see Table 1). The overall WFM system fine resolution, about $4.3' \times 4.3'$, is achieved by combining two cameras (orthogonal directions) composing 1 unit (Fig. 1, bottom panel). The

E-mail: immacolata.donnarumma@iaps.inaf.it, Telephone: +39-064993-4117

Table 1. The LOFT WFM scientific requirements

| Parameter | 1 Camera | 1 Unit | Overall |
|---|---|---|---|
| Energy range | 2–50 keV | 2–50 keV | 2–50 keV |
| | (extended to 80 keV) | (extended to 80 keV) | (extended to 80 keV) |
| Geometric Area (cm$^2$) | 182 | 364 | 1820 |
| Peak effective area (cm$^2$) | $> 40$ | $> 80$ | $> 80$ |
| Energy resolution | $< 300$ eV | $< 300$ eV | $< 300$ eV |
| Field of View at zero response | $90° \times 90°$ | $90° \times 90°$ | $180° \times 90° + 90° \times 90°$ |
| Angular Resolution | $5' \times 5°$ | $5' \times 5'$ | $5' \times 5'$ |
| Point source location accuracy (10-$\sigma$) | $1' \times 30'$ | $1' \times 1'$ | $1' \times 1'$ |
| On-axis 5-$\sigma$ sens. in 3 s (Gal Centre) | 380 mCrab | 270 mCrab | 270 mCrab |
| On-axis 5-$\sigma$ sens. in 50 ks (1-day Gal Centre) | 3 mCrab | 2 mCrab | 2 mCrab |

overall configuration of the WFM[7] envisages a set of 5 units (Fig. 1, top panel), for a total geometric area of 1820 cm$^2$, located on the top of the satellite tower. The units are off-set between each other along one direction in order to provide the maximum coverage of the sky region accessible to the LAD ($\sim 4$ sr). Moreover, the unit FoVs are partially overlapped, providing a good system redundancy.

In order to increase the fully coded field of view (and then to optimize the sensitivity at large off-axis angles) each camera has a Tungsten mask ($\sim 260 \times 260$ mm$^2$) about 1.8 larger than the detector. X-ray photons outside the field of view are shielded by the collimator which also acts as the mechanical structure supporting the mask. The scientific requirements of the WFM are summarized in Table 1.

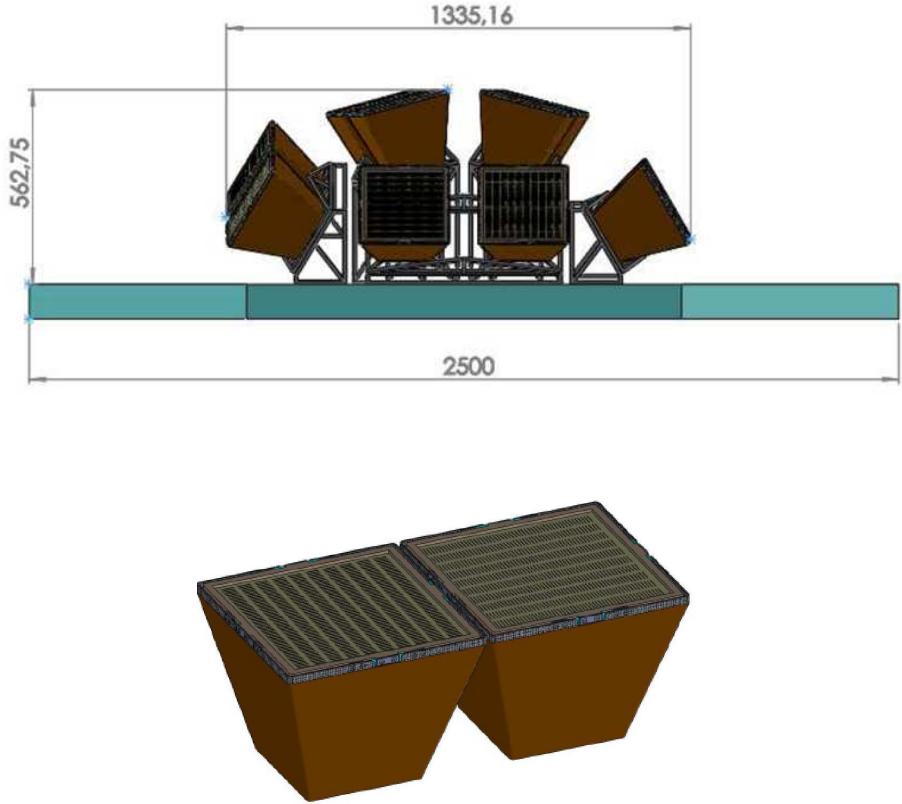

Figure 1. Top panel: 5 units placement, as seen from nominal anti-Sun direction. Bottom panel: two cameras forming one WFM unit, oriented orthogonally to each other.

## 2. SIMULATOR

In order to estimate the imaging performance of the SDD-based WFM system, we developed an end-to-end model able to simulate the system response to point sources and to Cosmic X-ray Background (CXB). The development was based on both Monte Carlo and analytical approaches to enable the simulations to be as realistic and quick as possible. The main steps can be summarized as follows:

- generation of a photon list containing arrival time, energy, and direction of the photons (for both point sources and diffuse background) for a certain pointing;

- propagation of the photons through the mask pattern;

- propagation of the photons from the mask into the detector body;

- application of the SDD spatial and spectral resolutions.

We will give in the following sections an overview of the WFM simulator and of the source reconstruction capabilities. It is worth noticing that the simulator deals with 1 camera at a time and that we will show the performance of two cameras composing one WFM unit.

### 2.1 The mask code

The mask pattern consists of 1040 × 16 elements (fine and coarse directions, respectively; see Fig. 2, left panel) distributed following a bi-quadratic residue method (Uniformly Redundant Array, prime number equal to 16901) with a resulting open fraction of 25%, which we consider as the best trade-off for the observations of bright and faint sources. Indeed, it can be considered optimized for fainter sources,[8] if the internal background (introduced e.g. by particles or electronic noise) is negligible with respect to the external one (X-ray diffuse + other sources). This is the case for the LOFT WFM, that is dominated by the external background due to the wide field of view: the internal background (mainly particles) is only 15% of the external one (see [9]). Moreover, another important advantage of selecting a 25% open fraction mask for the LOFT WFM is the telemetry saving.

We studied the mask properties as a function of the off-axis angle for both the fine and the coarse resolution directions. In particular, we estimated the mask vignetting as a function of the off-axis angle considering the distribution in the mask code of the open elements, and the probability that two or more open elements can occur one after the other. We note that vignetting strongly depends on the ratio between the mask element size and the mask thickness. Since the mask element size is 250 $\mu$m and 16.125 mm for the fine and coarse direction, respectively and the mask thickness is 150 $\mu$m, imaging capabilities in the coarse directions will be hardly affected by vignetting in the required WFM FoV (Table 1). We show in the right panel of Fig. 2 the variation of the mask open fraction across the FoV due to our choice of the mask pattern for both fine and coarse directions (black and red lines, respectively).

### 2.2 The detector model

The detector model used in the imaging simulator is composed by 4 SDD tiles (see Fig. 3, left panel), each with a sensitive area of 6.51 × 7.0 cm$^2$ (7.74 × 7.25 cm$^2$ geometric area) and a thickness of 450 $\mu$m. A 6.15 mm wide guard zone surrounds two sides of each tile, thus creating a cross-shaped non-sensitive area of the detector, 12.30 mm long in one dimension and 2.5 mm long in the other. The definition of the Silicon sensitive area includes the absorption layers on the detector surface, the undepleted regions between the drift cathodes and the fully-depleted Silicon bulk. We also consider a small gap (100 $\mu$m) between the tiles in order to account for the space needed during the plane assembly procedure. This results in a ratio between geometric and sensitive area of about 90%.

The absorption layers on top of the SDD (sketched in the right panel of Fig. 3) are the following:

- Passivation: Low Temperature Oxide (LTO) SiO$_2$ (0.54 $\mu$m thick, 100% of the surface)

- Metal Contact: Al (0.5 $\mu$m thick, 93.3% of the surface)

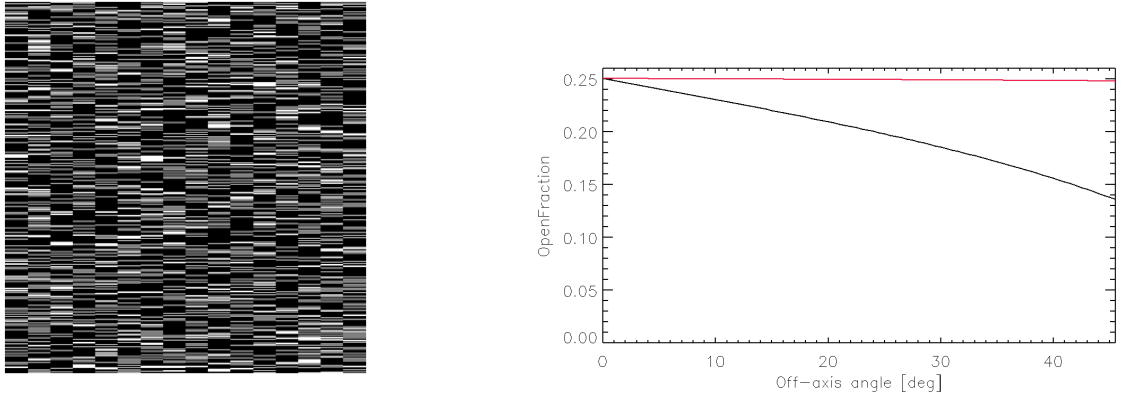

Figure 2. Left panel: mask pattern of biquadratic residues type; coarse direction (horizontal), fine direction (vertical). Right panel: mask open fraction as a function of off-axis angle. Black line: fine resolution direction; red line: coarse resolution direction.

- Field Oxide: $SiO_2$ (0.16 $\mu$m thick, $\sim 100\%$ of the surface)
- Undepleted p implant: Si (0.5 $\mu$m thick, 0.83% of the surface)
- Undepleted Si bulk: ($\sim 4$ $\mu$m thick, 16.7% of the surface)

The physical model of the detector takes into account the real structure of the SDD. It is worth noticing that the detector does not have any surface structure which physically separates the anodes: the SDD can be thus considered as a continuous, non-pixelated detector, while along the drift direction the cathode implants are repeated every 120 $\mu$m (which corresponds to about 1/135 the coarse mask pitch).

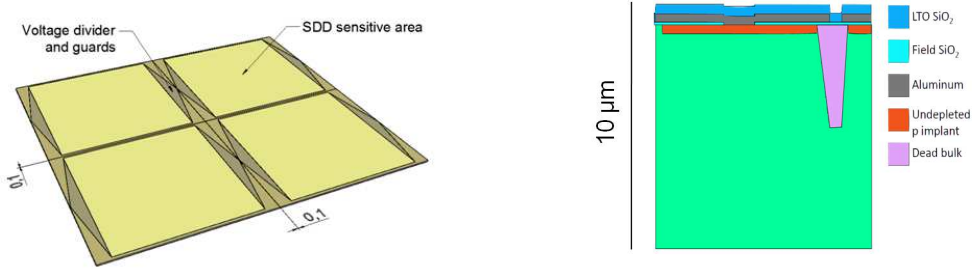

Figure 3. left panel: sketch of the detector plane. Right panel: passive layers on the detector surface; the plot shows only the first 10 $\mu$m of the detector.

### 2.3 Photon propagation through the mask

The first step of the photon list processing is the mask crossing which includes both the transparency and the vignetting (see section 2.1). The propagation of the photons through the mask first requires that each incoming photon direction is converted from the equatorial coordinates to the reference system of the mask, taking into account the pointing direction (right ascension, declination and roll angle). The impact points of photons on the mask plane (top layer) are obtained by using a uniform distribution random generator and by accounting for photon directions in the mask reference system. The projections of the impact points on the bottom boundary of the mask are also obtained in order to calculate the path length of the photons inside the Tungsten and to determine the probability of absorption. A random number generator decides whether a particular photon is absorbed or not based on this probability. As a final step, a new photon list is produced, that includes the photon directions, the impact point in the mask reference frame and a flag which defines if photons crossed the mask or not.

## 2.4 Photon propagation from the mask to the detector frame

The photon list resulting from the previous step is provided to another procedure that calculates the photon propagation from the mask to the detector. Considering the mask-detector system geometry (mask and detector sizes, mask-detector distance), photons are projected on the detector frame, it is verified if these fall in the sensitive area and whether they are absorbed within the dead layers or in the Silicon depleted bulk. Then, for the photons not absorbed in the dead layers, we obtain the coordinates $(x, y, z)$ in the Silicon bulk corresponding to the actual position where they are absorbed due to photoelectric effect. This allows to account for the different focal length experienced by photons of different energies, i.e. to consider the effect introduced by the inclined penetration which may result in a different (x,y) position than if they were detected at the top layers of the detector. In Fig. 4 we report the average absorption depth in the Silicon bulk as a function of the energy.

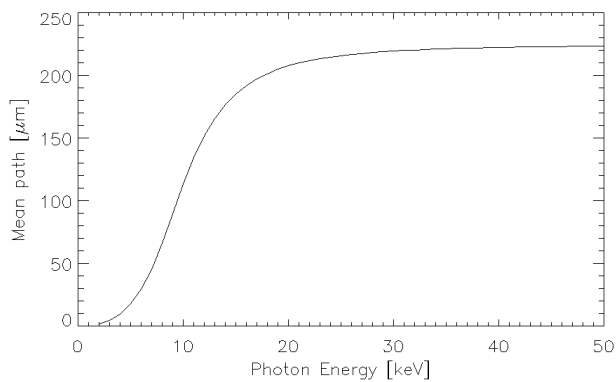

Figure 4. Mean absorption depth in the Silicon bulk (450 $\mu$m thick) as a function of the energy.

## 2.5 Application of the SDD spatial and spectral resolutions

Starting from the coordinates of the photon absorption point, and taking into account the photon energy, we apply to the overall photon list the SDD spatial and spectral resolution. As described in [4], the electron cloud generated by photon interaction with the detector can be sampled by one or more anodes and, moreover, the signal will be affected by a coherently varying baseline due to the common-mode noise. The energy resolution, therefore, deteriorates with the number of anodes over which the signal spreads, and improves with the number of channels used to evaluate the common-mode noise. In order to account for the variation of the energy resolution as a function of the photon impact point, energy-dependent maps have been generated and given as an input to the simulator, providing the photon energy resolution on the basis of the simulated photon impact point. Fig. 5 shows the spectral resolution of a 2 keV photon as a function of the photon absorption point $(x_0, y_0)$, for a detector with 145 $\mu$m pitch at $-20°$C. The variation of the SDD spatial resolution along the drift and anodic directions as a function of photon impact point is also properly considered. We refer to [5] for details about spatial resolution maps.

## 3. THE EXPECTED IMAGING PERFORMANCE

As for the imaging capabilities, the SDD is a continuous detector for which it is possible to apply a finely sampled cross-correlation procedure in order to improve the image reconstruction.[10] This means that both the detector image and the reconstruction matrix are finely sampled: they contain $r > 1$ samples per resolution element. In detail, a value of $r = 16$ in fine and coarse directions (which corresponds to a detector element size of $\sim 16$ $\mu$m $\times 1$ mm) allows us to study the systematic effects affecting the point spread function, e.g. those due to the inclined penetration and the mask vignetting. However, it is worth noticing that this fine sampling could represent an issue for the computational power, as long as detector image accumulation and/or deconvolution procedure are performed on-board. Indeed, a value of $r \sim 3$ should allow to sample most of the spatial systematics in the image during in-flight operations. In our simulations, we have adopted a value of $r = 16$ to better characterize the point spread function within the field of view, while a smaller value ($r = 8$) is assumed for the other imaging simulations presented below. We performed a set of simulations of an isolated Crab-like source, located in different places in the WFM camera FoV in order to estimate the point source sensitivity of the WFM system. Finally, we will show the simulation of 10 ks exposure of the Galactic bulge, aimed at showing the performance of our system in crowded fields.

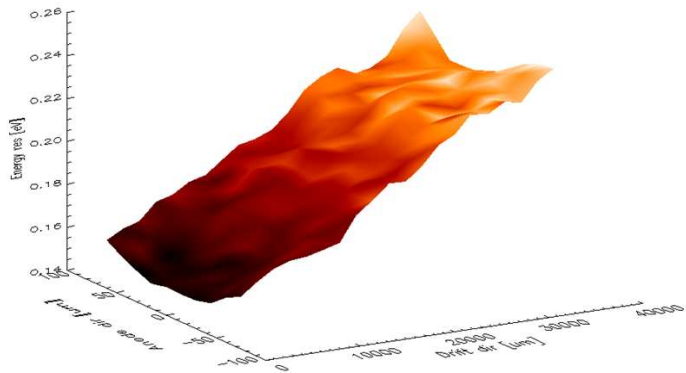

Figure 5. Energy resolution map at 2 keV as a function of photon impact point.

### 3.1 The system PSF

We studied the system Point Spread Function (PSF) as a function of the photon energy and of the source off-axis angle. A set of simulations of the Crab-like source at different positions in the WFM FoV has been performed. These are summarized in Table 2, where $\theta_x$ and $\theta_y$ represent the angular displacement in the coarse and fine directions, respectively. All the simulations contain a 1 ks exposure of the Crab Nebula plus the Cosmic X-ray background. For the Crab Nebula, we considered an absorbed power-law spectrum with $\Gamma = 2.1$, $N_H = 4 \times 10^{21}$ cm$^{-2}$ and a normalization of 9.5 ph cm$^{-2}$ s$^{-1}$ keV$^{-1}$ at 1 keV, while we adopted the Gruber model [11] for the X-ray background.

Figure 6 shows the 2–20 keV PSF along the coarse resolution direction (left panels) and the fine resolution direction (right panels) for the Crab at $0°$ and $-30°$ off-axis in the fine direction (on-axis in the coarse direction). We found that the PSF is triangular shaped with a FWHM in the two directions of $4.3' \times 4.8°$ (see Fig. 6), and remains almost constant when the off-axis angle is small (<10–20 degrees). For larger off-axis angles, the inclined penetration and the vignetting become important as reported in section 2 and the PSF shape changes accordingly. In Fig. 7 (right panel) we report the case of the Crab at $\theta_x = 0°$, $\theta_y = -30°$. Right panels show the PSF in four energy bands (2–4 keV, 4–6 keV, 6–10 keV and 10–20 keV) for the fine resolution direction. The effect of vignetting is visible at low energies (first and second right panel) as a flattening of the PSF peak. Instead, the inclined penetration becomes more important as the photon energy increases (see 3rd and 4th right panels of Fig. 7). The penetration effect can be considered primarily as a "blurring" of the PSF, due to the fact that the dependence of the photon absorption depth with the energy translates in an energy dependence of the mask-detector distance. Moreover the peak position changes up to $\sim 1$ arcmin between 2 and 20 keV, adding a systematic off-set in the point source location accuracy for off-axis sources when the full energy band is considered in the imaging procedure. Indeed, this can be properly included in the PSF modeling performed to assess flux and position of the sources. In this regard, pre-flight calibrations of the spatial response will play a key role.

These simulations provide the point source 5-$\sigma$ sensitivity per camera resulting from a 1-day exposure (50 ks), which is reported in Table 2 expressed in mCrab. The single camera daily sensitivity is of the order of 3 mCrab for on-axis isolated source. The sensitivity of one unit improves by a factor of $\sqrt{2}$ for sources with known position. It is worth noticing that for the localization of new sources we require an independent detection by both cameras, and then we will be limited by the sensitivity of the single camera. However, once the source is firmly detected, the full unit can be used for its spectral analysis, exploiting the full effective area and sensitivity.

Table 2. Single camera point source sensitivity in the WFM FoV (as a function of the coarse and fine directions). In the 3rd column the signal to noise ratio obtained in 1 ks for a Crab-like source is reported.

| $\theta_x$ (degrees) | $\theta_y$ (degrees) | SNR | 1 Camera Sensitivity (50 ks) (mCrab) |
|---|---|---|---|
| 0 | 0 | 225.2 | 3.1 |
| -10 | 0 | 220.2 | 3.2 |
| 0 | -10 | 209.0 | 3.4 |
| -30 | 0 | 107.0 | 5.6 |
| 0 | -30 | 108.7 | 6.5 |

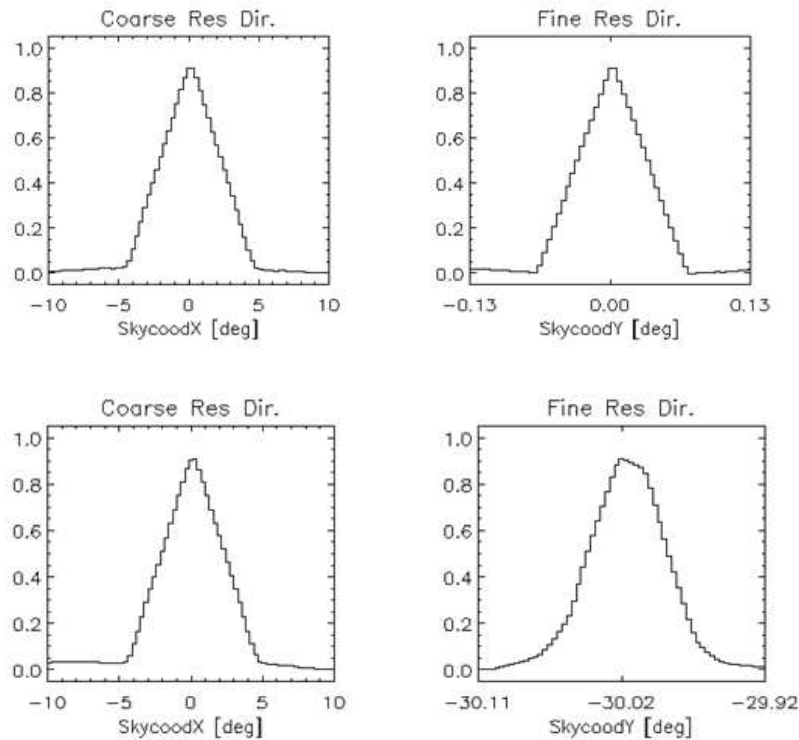

Figure 6. 2–20 keV system point spread function along the coarse resolution direction (left) and along the fine resolution direction (right) for simulations of the Crab Nebula (0°, −30° off-axis in the fine direction).

## 3.2 Galactic bulge

In order to study the instrument imaging response to crowded fields, we simulated a 10 ks observation of the Galactic bulge by using a photon list generated from the 4th INTEGRAL/IBIS catalogue (see [12]). The catalogue contains 723 sources detected by the IBIS gamma-ray imager in the energy band 20–100 keV. Although the IBIS energy range only partially overlaps the LOFT WFM energy band, it has been adopted because it represents one of the deepest survey in hard X-rays. The simulated field takes into account the source flux measured by INTEGRAL/IBIS in the 20–40 keV energy band, properly rescaled to the WFM primary energy range (2–50 keV).

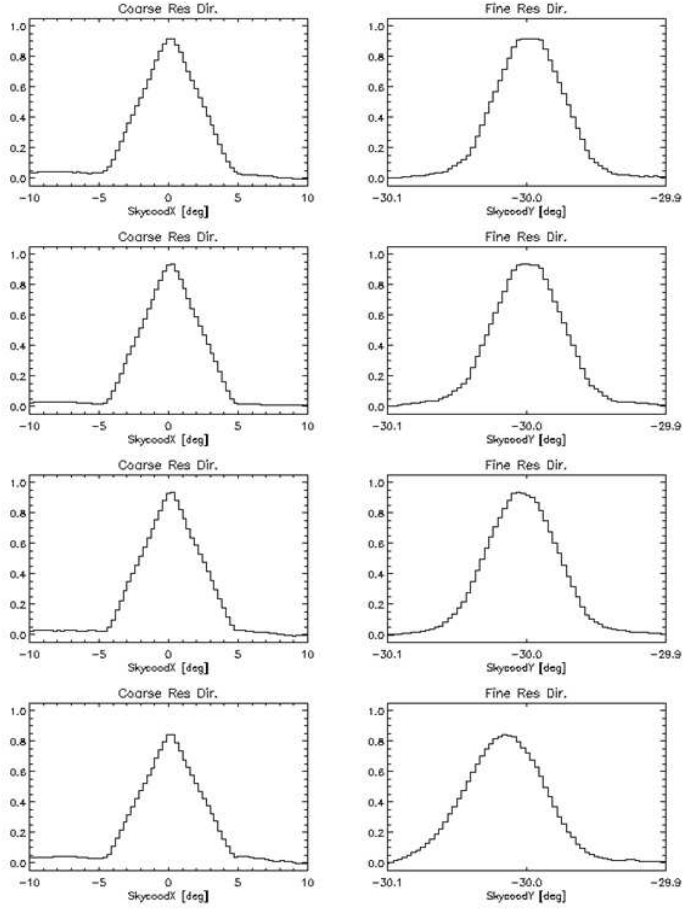

Figure 7. Right panel: PSF for a Crab like spectrum in four energy bins. The effects of the vignetting (flattening of the PSF) and of the inclined penetration (blurring of the PSF and drift of the source peak as a function of energy) are clearly visible.

In Fig. 8 (top panel), we report the two camera images of the Galactic bulge expressed in terms of source significance. More than 20 sources (over $\sim 80$ with a flux greater than a few mCrab) are clearly detected with a signal to noise ratio $> 15$ (10 ks exposure). The strips in the Galactic bulge view of each camera show the angular resolution of 1 single camera. The fine imaging of each WFM unit is achieved by combining the two images, as shown in Fig. 8 (bottom panel). In this representation, we superimposed to the image the source name and a circular region which is about 2 times larger than the fine angular resolution ($4.3' \times 4.3'$). The sources reported in Fig. 8 spans a flux range from a few tens up to a few hundreds of mCrab, with a signal to noise ratio between 18 and 180. We note that Sco X-1, the dominant source of photons in the Galactic bulge, has been excluded by the simulations in order to limit the coding noise. This means that the image sensitivity should be worse by a few tens of percent.

## 3.3 Point Source Location Accuracy

In the case of the Galactic bulge, we performed a catalogue search in order to estimate the point source location error in both the coarse and fine resolution directions of a single camera sky image ($\theta_x$, $\theta_y$). Fig. 9 shows the resulting distribution of $\theta_x$ and $\theta_y$ as a function of the source signal to noise ratio. The reasonably good Point Source Location Accuracy in the coarse direction (in addition to the excellent one in the fine) highlights the benefit of having two independent cameras: even in the case of failure of one camera the other one will still deliver bidimensional images of the same field, although with a coarser resolution in one direction. This configuration offers an intrinsic redundancy to the system. We combined the Galactic bulge simulation with those of the Crab Nebula in order to estimate the system Point Source Location Accuracy (PSLA) as a function of the source significance when the information provided by two orthogonal cameras are used simultaneously. The PSLA can be expressed as:

$$PSLA \sim \theta/SNR \qquad (1)$$

where $\theta$ is the system angular resolution and SNR the signal to noise ratio. For each detected source, we have performed a PSF fitting procedure with a triangular function plus a constant (to account for the image baseline). The fitting is simultaneously carried out on the two orthogonal sky images. Fig. 9 represents the PSLA as a function of the SNR obtained by performing the squared sum of the source significances in the two cameras (as remarked before, this applies for known sources). The red line represents the expected PSLA for one WFM unit, that is $\sim 0.9'$ at 10 $\sigma$ (for comparison the PSLA in coarse direction is $< 50$ arcmin). The results obtained with a simple PSF fitting demonstrate that the system fulfills the requirement of a point source location accuracy better than 1'. Anyway, it is worth noticing that the PSLA estimated by means of a PSF fitting procedure cannot take into account all the physical effects discussed in Section 3.1 (inclined penetration, mask vignetting, etc.) and their dependence on the energy and off-axis position. As previously discussed, these effects introduce a systematic error in the positioning accuracy causing a deviation of the real PSLA distribution with respect to the ideal one. The PSLA is then expected to improve with a more refined PSF fitting procedure, able to account for all the possible systematic effects affecting its shape.

## 4. CONCLUSIONS

The WFM aboard LOFT is composed of 5 units, each of them composed by two cameras which are orthogonally oriented to each other. We have presented the end-to-end model simulator we developed in order to estimate sensitivity and imaging capabilities of each unit. It works with 1 camera per time but simulations in orthogonal directions have been performed in order to estimate the imaging performance of the overall system. Each camera has asymmetric imaging capability, due to the different spatial resolution in anodic and drift directions in SDDs, resulting in a few arcmin × a few degrees positioning of sources in the two directions. Although angular resolution is coarse in one direction, the actual configuration offers a good redundancy of the system in case of a camera failure.

We developed a Monte Carlo + analytical end-to-end model able to simulate typical astrophysical observations considering both source and background. All the known physical effects related to the mask-detector system have been included in the simulations. In particular, the mask transparency and vignetting affecting the imaging properties at larger off-axis angles are well accounted for: simulations performed for Crab-like sources at different off-axis angles provide results in agreement with the expectations based on the mask pattern code and geometry. Photon interaction in SDDs takes into account the dependence of both energy and spatial resolution on the photon impact point. Inclined penetration in the detector bulk is also considered. All these effects are expected to significantly affect the photon position and reconstruction if not properly taken into account. In this regard, we have studied the behavior of the PSF of one camera as a function of energy and position across the FoV, by simulating a Crab-like (isolated) point source. We find that its shape clearly reflects the deviation from a purely triangular shape expected for a finely sampled detector as the off-axis angle increases. In particular, the PSF behavior as a function of energy shows the blurring due to both vignetting and inclined penetration into the detector. In addition, if we look at the PSF integrated over the source spectrum, the energy dependent blurring results in a shift of 1 arcmin in the PSF peak at 30° off-axis in the fine direction. This is likely associated to the average effect provided by the different focal lengths experienced by photons with different energy. The characterization of the PSF enabled to study both the sensitivity of our system and the point source location accuracy of 1 unit (2 orthogonal cameras). In particular, we have estimated the 1-day 5-$\sigma$ sensitivity for 1 unit which is about 2 mCrab for on-axis isolated sources. In addition, we have simulated sources in the Galactic bulge to give a preliminary estimate of the sensitivity in crowded fields. We found that sources with flux greater than

15 mCrab are visible in 10 ks, but we cannot do any statement about 1 unit WFM sensitivity based on these numbers, since the coding noise variance of the combined images still includes the coding noise introduced by all the sources in the FoV. The Iterative Removal of Sources (see [13,14]) has to be performed in order to obtain cleaned images and then the real sensitivity in crowded fields, which is expected to improve with respect to the number obtained so far. This work is on going.

Our preliminary results show that the SDD-based WFM is able to fulfill the LOFT mission scientific requirements reported in Table 1.

## REFERENCES

[1] Feroci, M., Stella, L., van der Klis, M., Courvoisier, T. J.-L., Hernanz, M., Hudec, R., Santangelo, A., Walton, D., Zdziarski, A., Barret, D., Belloni, T., Braga, J., Brandt, S., Budtz-Jørgensen, C., Campana, S., den Herder, J.-W., Huovelin, J., Israel, G. L., Pohl, M., Ray, P., Vacchi, A., Zane, S., Argan, A., Attinà, P., Bertuccio, G., Bozzo, E., Campana, R., Chakrabarty, D., Costa, E., de Rosa, A., Del Monte, E., di Cosimo, S., Donnarumma, I., Evangelista, Y., Haas, D., Jonker, P., Korpela, S., Labanti, C., Malcovati, P., Mignani, R., Muleri, F., Rapisarda, M., Rashevsky, A., Rea, N., Rubini, A., Tenzer, C., Wilson-Hodge, C., Winter, B., Wood, K., Zampa, G., Zampa, N., Abramowicz, M. A., Alpar, M. A., Altamirano, D., Alvarez, J. M., Amati, L., Amoros, et al., "The Large Observatory for X-ray Timing (LOFT)," *Experimental Astronomy* , 100 (Aug. 2011).

[2] Feroci, M., Costa, E., Soffitta, P., Del Monte, E., di Persio, G., Donnarumma, I., Evangelista, Y., Frutti, M., Lapshov, I., Lazzarotto, F., Mastropietro, M., Morelli, E., Pacciani, L., Porrovecchio, G., Rapisarda, M., Rubini, A., Tavani, M., and Argan, A., "SuperAGILE: The hard X-ray imager for the AGILE space mission," *Nuclear Instruments and Methods in Physics Research A* **581**, 728–754 (Nov. 2007).

[3] Tavani, M., Barbiellini, G., Argan, A., Bulgarelli, A., Caraveo, P., Chen, A., Cocco, V., Costa, E., de Paris, G., Del Monte, E., Di Cocco, G., Donnarumma, I., Feroci, M., Fiorini, M., Froysland, T., Fuschino, F., Galli, M., Gianotti, F., Giuliani, A., Evangelista, Y., Labanti, C., Lapshov, I., Lazzarotto, F., Lipari, P., Longo, F., Marisaldi, M., Mastropietro, M., Mauri, F., Mereghetti, S., Morelli, E., Morselli, A., Pacciani, L., Pellizzoni, A., Perotti, F., Picozza, P., Pontoni, C., Porrovecchio, G., Prest, M., Pucella, G., Rapisarda, M., Rossi, E., Rubini, A., Soffitta, P., Trifoglio, M., Trois, A., Vallazza, E., Vercellone, S., Zambra, A., Zanello, D., Giommi, P., Antonelli, A., and Pittori, C., "The AGILE space mission," *Nuclear Instruments and Methods in Physics Research A* **588**, 52–62 (Apr. 2008).

[4] Campana, R., Zampa, G., Feroci, M., Vacchi, A., Bonvicini, V., Del Monte, E., Evangelista, Y., Fuschino, F., Labanti, C., Marisaldi, M., Muleri, F., Pacciani, L., Rapisarda, M., Rashevsky, A., Rubini, A., Soffitta, P., Zampa, N., Baldazzi, G., Costa, E., Donnarumma, I., Grassi, M., Lazzarotto, F., Malcovati, P., Mastropietro, M., Morelli, E., and Picolli, L., "Imaging performance of a large-area Silicon Drift Detector for X-ray astronomy," *Nuclear Instruments and Methods in Physics Research A* **633**, 22–30 (Mar. 2011).

[5] Evangelista, Y. et al. *in SPIE Conf. Ser., 8443-210* (2012).

[6] Skinner, G. K., "Sensitivity of coded mask telescopes," Appl. Opt. **47**, 2739–2749 (May 2008).

[7] Brandt, S. et al. *in SPIE Conf. Ser., 8443-88* (2012).

[8] in 't Zand, J. J. M., Heise, J., and Jager, R., "The optimum open fraction of coded apertures. With an application to the wide field X-ray cameras of SAX," A&A **288**, 665–674 (Aug. 1994).

[9] Campana, R. et al. *in SPIE Conf. Ser., 8443-209* (2012).

[10] Fenimore, E. E. and Cannon, T. M., "Uniformly redundant arrays - Digital reconstruction methods," Appl. Opt. **20**, 1858–1864 (May 1981).

[11] Gruber, D. E., Matteson, J. L., Peterson, L. E., and Jung, G. V., "The Spectrum of Diffuse Cosmic Hard X-Rays Measured with HEAO 1," *The Astrophysical Journal* **520**, 124–129 (1999).

[12] Bird, A. J., Bazzano, A., Bassani, L., Capitanio, F., Fiocchi, M., Hill, A. B., Malizia, A., McBride, V. A., Scaringi, S., Sguera, V., Stephen, J. B., Ubertini, P., Dean, A. J., Lebrun, F., Terrier, R., Renaud, M., Mattana, F., Götz, D., Rodriguez, J., Belanger, G., Walter, R., and Winkler, C., "The Fourth IBIS/ISGRI Soft Gamma-ray Survey Catalog," ApJS **186**, 1–9 (Jan. 2010).

[13] Hammersley, A. P., PhD thesis, University of Birmingham (1986).

[14] in' t Zand, J., PhD thesis, Utrecht University, The Netherlands (1992).

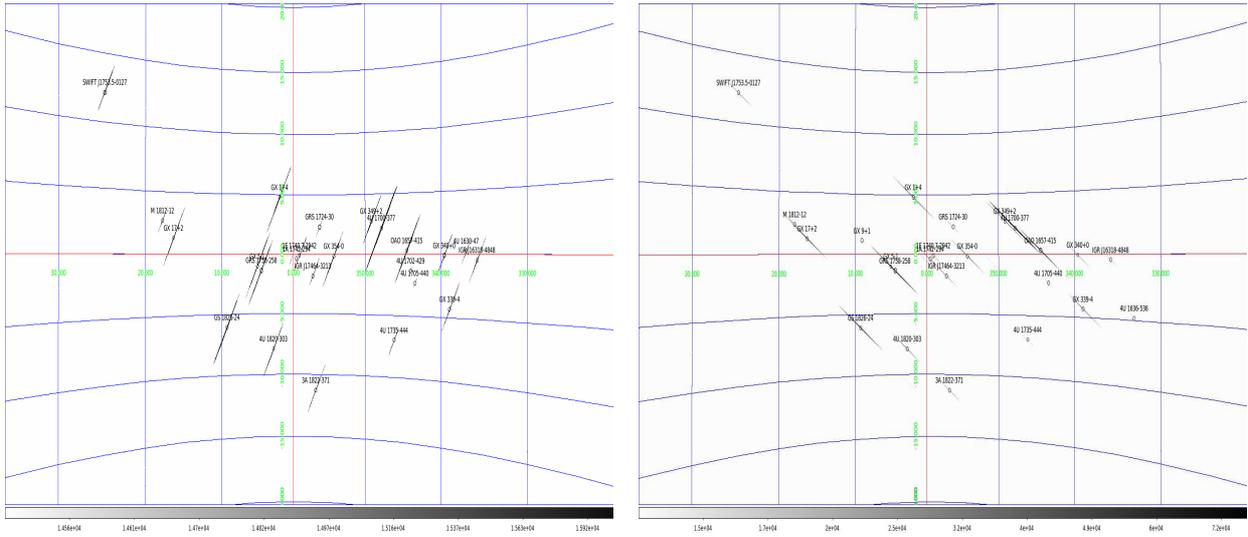
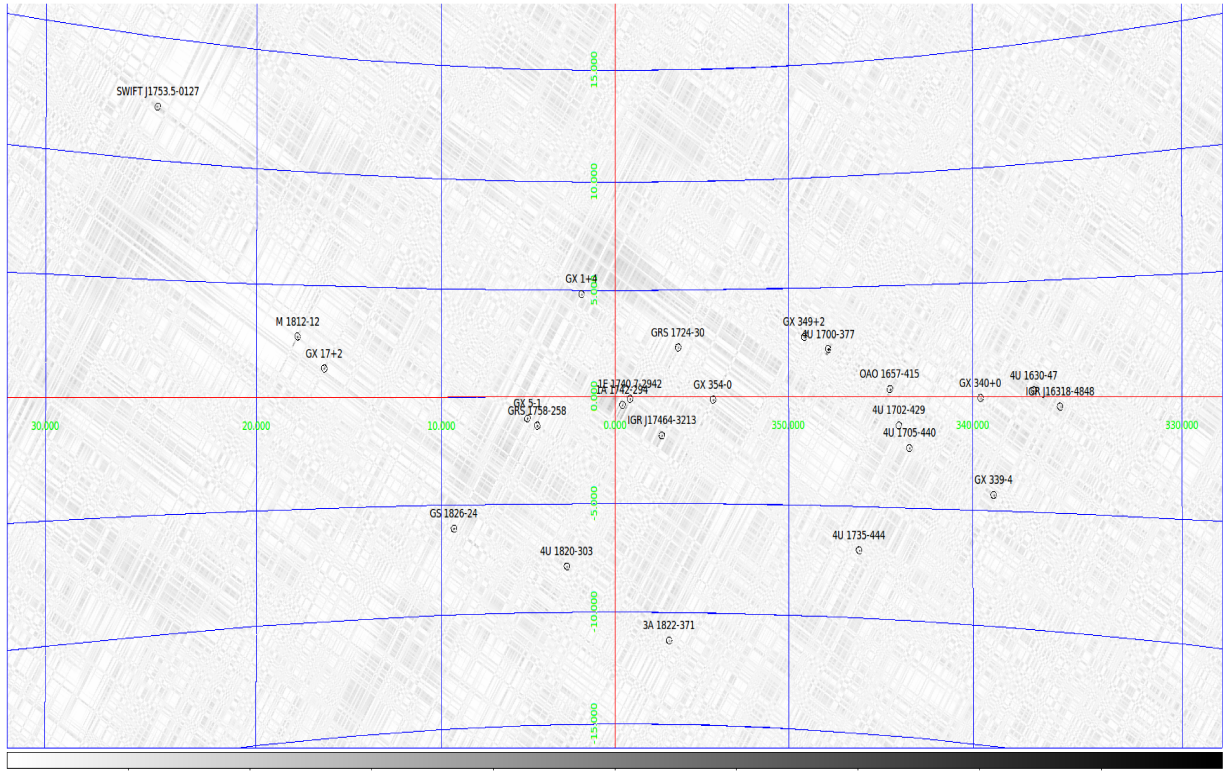

Figure 8. Top panel: 10 ks Galactic bulge simulation, the view of two single cameras. The strips in each plot show the angular resolution of each camera. Bottom panel: the fine imaging of each WFM Unit, achieved by combining the two images.

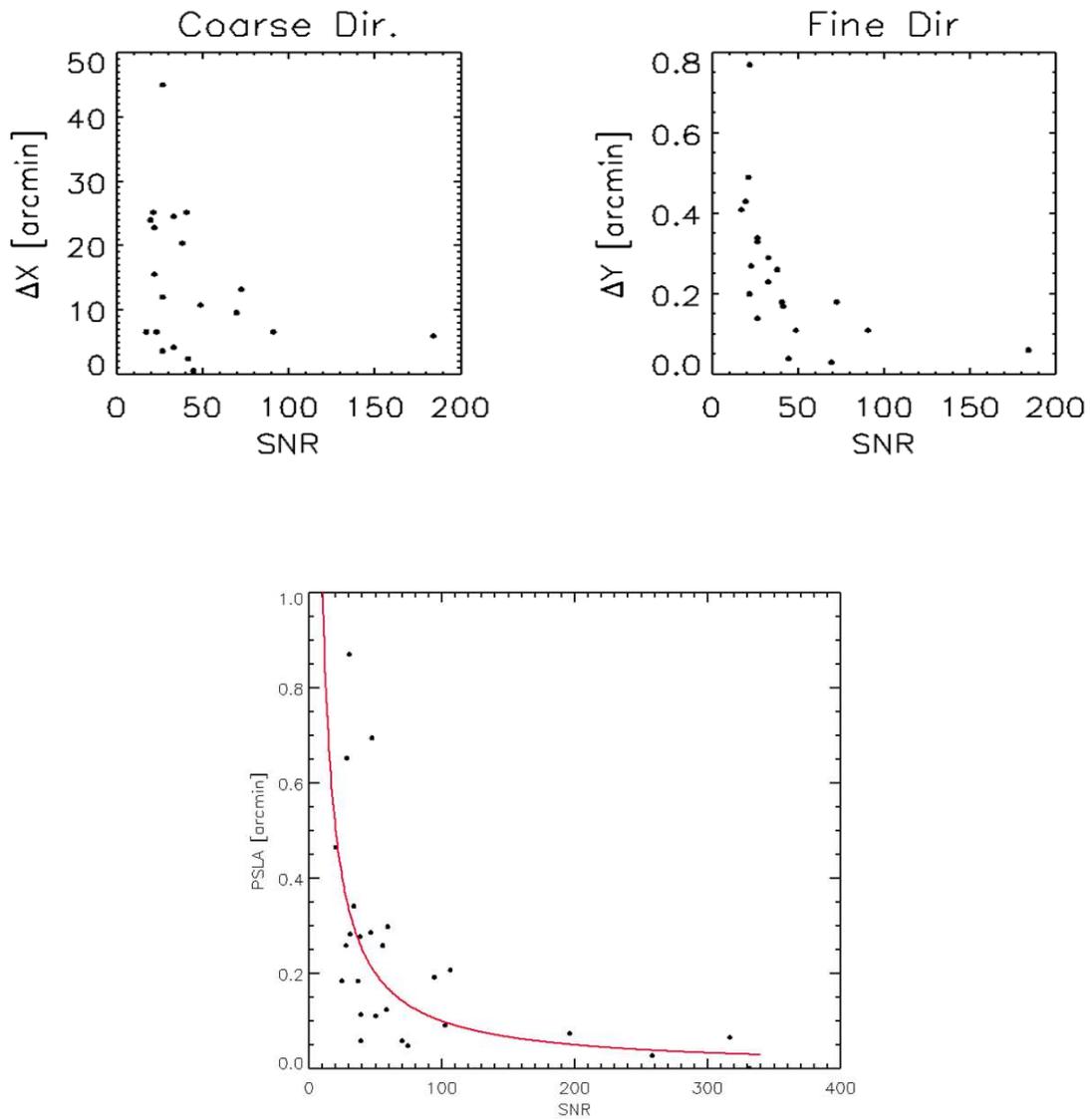

Figure 9. Top panel: Point Source Location Error as a function of the source SNR for a simulated 10 ks observation of the Galactic bulge in the coarse resolution direction (left panel) and in the fine resolution direction (right panel). Lower panel: Point Source Location Accuracy obtained by fitting a triangular function to the source peaks in the images. The red line represents the ideal PSLA assuming the system fine resolution.